\documentclass[structabstract]{aa}    
\usepackage{graphicx}
\usepackage{txfonts}
\newcommand{\itbf}[1]{\textbf{\textit{#1}}}
\newcommand{\fermi}[2]{\it Fermi}

\begin{document}
   \title{\itbf{INTEGRAL}  observations of the GeV blazar PKS~1502+106 and of the hard X-ray bright Seyfert galaxy Mkn~841\thanks{Based on observations made  with:  {\it INTEGRAL}, an ESA project with instruments and Science Data Centre funded by ESA member states (especially the PI countries: Denmark, France, Germany, Italy, Switzerland, Spain), Czech Republic and Poland, and with the participation of Russia and the USA,   and the Nordic Optical Telescope, operated on the island of La Palma jointly by Denmark, Finland, Iceland, Norway, and Sweden, in the Spanish Observatorio del Roque de los Muchachos of the Instituto de Astrofisica de Canarias.}}


   \author{E. Pian
          \inst{1,2,3}
          \and
          P. Ubertini\inst{4}
          \and
          A. Bazzano\inst{4}
          \and
          V. Beckmann\inst{5}
          \and
          D. Eckert\inst{6}
          \and
          G. Ghisellini\inst{7}
          \and
          T. Pursimo\inst{8}
          \and
          G. Tagliaferri\inst{7}
          \and
          F. Tavecchio\inst{7}
          \and
          M. T\"urler\inst{9}
          \and 
          S. Bianchi\inst{10}
                    \and 
          V. Bianchin\inst{11}
          \and
          R. Hudec\inst{12,13}
          \and
          L.  Maraschi\inst{7}
          \and
          C.M. Raiteri\inst{14}
          \and
          S. Soldi\inst{15}
          \and
          A. Treves\inst{16}
          \and
          M. Villata\inst{14}
}
   \institute{
    INAF, Trieste Astronomical Observatory, Via G.B. Tiepolo 11, 34143 Trieste, Italy\\
    \and
    Scuola Normale Superiore, Piazza dei Cavalieri 7, 56126 Pisa, Italy\\
    \and
    European Southern Observatory, Karl-Schwarzschild-Strasse 2
85748 Garching bei M\"unchen, Germany \\
                \email{elena.pian@sns.it}
    \and
    INAF, IASF-Roma, Via del Fosso del Cavaliere, 100, 00133 Roma, Italy\\
    \and
    Centre Fran\c{c}ois Arago, APC, Universit\'e Paris Diderot, CNRS/IN2P3, 10 rue Alice Domon et 
    L\'eonie Duquet, 75205 Paris Cedex 13, France \\
   \and 
   INAF/IASF Milano, Via E. Bassini 15, 20133 Milano, Italy\\
   \and
       INAF, Brera Astronomical Observatory, Via E. Bianchi 46, 23807 Merate (LC), Italy\\
    \and
    Nordic Optical Telescope, Apartado 474, 38700 Santa Cruz de la Palma, Spain\\
    \and
   {\it INTEGRAL} Science Data Centre, University of Geneva, Chemin d'Ecogia 16, 1290 Versoix, Switzerland\\    
   \and
   Department of Physics, University  "Roma Tre", Via della Vasca Navale 84, 00146 Roma, Italy \\
        \and
    INAF, IASF-Bologna, Via P. Gobetti, 40129 Bologna, Italy \\
    \and
    Astronomical Institute, Academy of Sciences, Fricova 298, 25165 Ondrejov, Czech Republic \\
        \and
Czech Technical University in Prague, Faculty of Electrical Engineering, Czech Republic \\
\and
  INAF,  Turin  Astronomical Observatory,  Via Osservatorio 20, 10025 Pino Torinese (TO),  Italy \\
        \and
        Laboratoire AIM - CNRS - CEA/DSM - Universit\'e  Paris Diderot (UMR 7158), CEA Saclay, DSM/IRFU/SAp, 91191 Gif-sur-Yvette, France
        \and
   Department of Physics and Mathematics, University of Insubria, Via Valleggio 11,  22100 Como,  Italy}

   \date{}

 
  \abstract
   {Extragalactic nuclear activity is best explored with observations at high energies, where the most extreme flux and spectral variations are expected to occur, witnessing changes in the accretion flow or in the kinematics of the plasma. 
In active galactic nuclei of blazar type, these variations are the most dramatic.}
   {By following blazar outbursts from their onset  and by correlating the observed variations at many different wavelengths we can reconstruct the behavior of the plasma and map out the development of the flare within the jet.  }
   {The advent of the {{\it Fermi}} satellite has allowed the start of a systematic and intensive monitoring program of blazars.   Blazar outbursts are very effectively detected by the LAT instrument in the MeV-GeV domain, and these can be promptly followed up with other facilities.  Based on a  {{\it Fermi}} LAT  detection of a high MeV-GeV state, we have observed the blazar PKS~1502+106 with the  {{\it INTEGRAL}}  satellite between 9 and 11 August 2008.  Simultaneous {{\it Swift}} observations have been also  accomplished, as well as optical follow-up at the Nordic Optical Telescope.}
   {The IBIS instrument  onboard  {{\it INTEGRAL}}  detected a source at a position inconsistent with the optical coordinates of PKS~1502+106, but consistent with those of the  Seyfert 1 galaxy Mkn~841, located  at 6.8 arcmin south-west from the blazar, which is therefore responsible for all the hard X-ray  flux detected by IBIS.  At the location of the blazar, IBIS sets an upper limit  of $\sim 10^{-11}$ erg~s$^{-1}$~cm$^{-2}$ on the 15-60 keV flux, that turns out to be consistent with a model of inverse Compton scattering accounting for the soft X-ray  and gamma-ray spectra measured by  {{\it Swift}} XRT and  {{\it Fermi}}  LAT, respectively.   The gamma-ray spectrum during the outburst  indicates substantial variability of the characteristic energy of the inverse Compton component in this blazar.      The hard X-ray state of the Seyfert appears to be nearly unchanged with respect to the past.  On the other hand, its soft X-ray flux (0.3-10 keV) varies  with a $\sim$50\% amplitude on  time scales from days to years.  On long time scales this is well correlated with the optical flux, with no measurable delay.}
   {In PKS~1502+106  the critical parameters that control variability are the accelerating power transferred to the relativistic electrons, and the magnetic field in the emitting region.  The spectrum of   Mkn~841 in the 0.3-100 keV range is well described by a power-law with a cutoff at $\sim$150 keV  and a Compton reflected continuum.}

   \keywords{galaxies: active -- quasars: individual:  PKS~1502+106 -- Galaxies: Seyfert -- galaxies: individual: Mkn~841 -- radiation mechanisms: general  -- gamma-rays: galaxies}
\authorrunning{E. Pian et al.}
\titlerunning{\itbf{INTEGRAL}  observations of  PKS~1502+106 and Mkn~841 }
   \maketitle
%

\section{Introduction}

The daily survey of the MeV-GeV sky by the Large Area Telescope (LAT) instrument (Atwood et al. 2009) onboard the {\it  Fermi} gamma-ray satellite  has enhanced dramatically the detection rate of flaring blazars, and will allow an accurate determination of their variability duty cycle at  the highest energies.  The follow-up of these high states at lower energies and the comparison with quiescent states make it possible to isolate the basic parameters that control the variations at all frequencies and ultimately to  set constraints on the behavior of the central engine (Ghisellini et al. 2009; Marscher et al. 2010; Abdo et al. 2010a; B\"ottcher \& Dermer 2010).

PKS~1502+106 ($z$ = 1.839, see Abdo et al., 2010b, for  a discussion on this preferred  redshift value)   is a bright, variable and optically polarized blazar (Impey \& Tapia 1988; George et al. 1994; Lahteenmaki \& Valtaoja 1999; Lister \& Homan 2005).  An et al. (2004) suggested, based on the detection of remarkable superluminal motion with high resolution radio observations, that it should be a strong gamma-ray emitter.   The source is not included in the Third  EGRET Catalog (Hartman et al. 1999), however it was detected in the first three months of sky-survey operation with the LAT (Abdo et al. 2009) and then again with the same instrument during a  flaring state on  6  August 2008 (Abdo et al. 2010b) with a flux above $2 \times 10^{-6}$ photons~cm$^{-2}$~s$^{-1}$ at energies larger than 100 MeV.  In January 2009, the source has undergone another flare reaching a comparable gamma-ray flux level (Abdo et al. 2010b).   It is one of the brightest,  most intrinsically luminous, and violently variable AGNs of the {\it  Fermi} LAT sample (Abdo et al. 2010c; Abdo et al. 2010d).

The field of  PKS~1502+106  had been observed on 24-26 January 2006 by {\it INTEGRAL} IBIS for a total integration time of 83 ks.  The  observation, analysed by Ubertini et al. (2008) in the light of the {\it  Fermi} report, resulted in the detection of a new source, IGR J15039+1022, at a
flux level of 1.6 mCrab (i.e., $2.45  \times 10^{-11}$ erg~cm$^{-2}$~s$^{-1}$ in the 18-60 keV range\footnote{Note that this flux supersedes the value of  $1.2  \times 10^{-11}$ erg~cm$^{-2}$~s$^{-1}$ reported previously (Ubertini et al. 2008; Abdo et al. 2010b).}).

The IBIS source, located  at RA = 15:03:53.76 and Dec = +10:22:08.4 (error radius of $\sim$5~arcmin), has been tentatively
identified with   the Seyfert~1 galaxy Mkn~841 ($z = 0.036$), and has been included in the Fourth {\it INTEGRAL} IBIS/ISGRI soft gamma-ray survey catalog (Bird et al. 2010) and in the second {\it INTEGRAL}  AGN catalog (Beckmann et al. 2009).      Ubertini et al. (2008) note that PKS~1502+106 is 11 arcmin away from IGR J15039+1022 and is not detected in the
January 2006 IBIS observation.   They derive a 2-sigma upper limit flux of 0.7 mCrab for the blazar
in the range 18-60 keV, assuming a Crab-like spectrum.       

Prompted by the {\it  Fermi} detection of the gamma-ray outburst, we activated {\it INTEGRAL} observations starting on 9 August 2008 (see preliminary images in Blazek \& Hudec 2009). Simultaneous observations with {\it  Swift} XRT and with the Nordic Optical Telescope (NOT) were also organized.  We report here the results and discuss them in the context of AGN models.


\section{Observations and data analysis}

\subsection{INTEGRAL}
The source PKS~1502+106  (whose catalogued optical  J2000 coordinates are RA = 15:04:24.98, Dec = +10:29:39.2) was observed by the {\it INTEGRAL} satellite (Winkler et al. 2003) for 203 ks in Revolution 711 between 2008 August 09, 01:53  UT and 2008 August 11,  15:12 UT.
The on-source time for IBIS/ISGRI  (Ubertini et al. 2003; Lebrun et al. 2003) was 194.5 ks.  The target is in the JEM-X  (Lund et al. 2003) field of view for 46 ks in  Rev. 711.

The screening, reduction, and analysis of the {\it INTEGRAL} IBIS/ISGRI  and JEM-X data have been 
performed using the {\it INTEGRAL} Offline Scientific Analysis (OSA) v.~7.0 and  v.~8.0, respectively,
publicly available through the {\it INTEGRAL} Science Data 
Center\footnote{http://www.isdc.unige.ch/INTEGRAL/download/osa\_sw} (ISDC,  Courvoisier et al. 2003).
The IBIS/ISGRI  and  JEM-X 
data have been accumulated into final images with algorithms  implemented in the software as described in Goldwurm et al.  (2003) and Westergaard et al. (2003), respectively.  Following similar methods, we have also re-analyzed  the archival {\it INTEGRAL} IBIS and JEM-X observation obtained on 24-26 January 2006 (Ubertini et al. 2008).  

The source has been also observed with the {\it INTEGRAL} OMC (Mas-Hesse et al. 2003) with a standard $V$-band Johnson filter.  The data have been extracted with default settings, using a $3 \times 3$ pixels binning, which is appropriate for point-like sources.  

\subsection{Swift}
X-ray and optical/UV observations of the blazar field have been accomplished from 7 to 22 August 2008 with {\it Swift} XRT and UVOT (Gehrels et al. 2004; Burrows et al. 2005;  Roming et al. 2005), respectively.  The pointing included the location of the Seyfert~1 Mkn~841.  The results of the analysis of the data of PKS~1502+106  have been reported by Abdo et al. (2010b).  Following similar procedures as those adopted by those authors, we have independently reduced  the XRT and UVOT data of the blazar and of the Seyfert~1.  
Like in Abdo et al. (2010b), for the blazar, given its low count rate   ($<0.5$ counts s$^{-1}$),  
only photon counting data  were considered and  XRT 
grades 0--12 were selected.  No pile-up correction was required.  For the Seyfert, we also considered window timing mode data.
The source events were extracted in circular regions centered on the sources 
with radii depending on the source intensity (10--20 pixels, 1 pixel $\sim2\farcs
37$),  while background events were extracted in source-free annular regions around the sources.

XRT spectra were extracted for each of the two sources from the observation of 8 August 2008,  that is the one with the longest exposure time  (12.5 ks).  
Ancillary response files were generated with {\sc xrtmkarf}  
and account for different extraction regions, vignetting and 
Point-Spread Function (PSF) corrections. 
We used the v010 spectral redistribution matrices  in the Calibration Database maintained by HEASARC.
All spectra were rebinned with a minimum of 20 counts per energy bin 
to allow $\chi^2$ fitting within {\sc XSPEC} (v11.3.2).   For the spectral fitting we fixed the hydrogen column density to the Galactic value of $2.22 \times 10^{20}$ cm$^{-2}$ (Kalberla et al. 2005). 

We extracted XRT light curves from the same image regions as for the spectra  
for source and background in the band 0.2--10\,keV.
For our analysis, we considered  time bins ranging from 1000 to 12000 s,   to ensure that each light curve point would have at least 
30 source-plus-background counts. 
The exposure requirements were that the bins be at least 50\,\% exposed.
In each case, the light curves were corrected for  
PSF losses, i.e., losses due to the extraction region geometry, 
and bad/hot pixels and columns falling within this region.

Likewise, the UVOT data of both sources were reduced and analyzed following procedures similar to those used by Abdo et al. (2010b).   The photometry of Mkn~841 may include partial contamination from the host galaxy bulge.  We used an aperture of 6$^{\prime\prime}$ for the v,b,u filter images and 12$^{\prime\prime}$ for the UV filters.  We have also reduced and analyzed the Mkn~841 data of 3 February 2007, that have been included in the Seyfert UVOT study of Grupe et al. (2010). 

\subsection{Nordic Optical Telescope}
We have observed the IBIS field on 13 August 2008 with the ALFOSC camera and optical BVRI filters at the NOT (Canary Islands).  The data have been reduced following standard procedures within the IRAF package (overscan subtraction, bias correction, twilight flat fielding).  Aperture photometry was performed with the IRAF routine APPHOT.
Eight field stars were used to estimate the photometric zero point
and their magnitudes were taken from the SDSS DR7 (using CDS/Aladin).
The field stars SDSS gri- magnitudes were converted to $BVRI$ bands following the transformations\footnote{http://www.sdss.org/DR5/algorithms/sdssUBVRITransform.html} recommended for the SDSS Data Release 5. See Table~1 for a log of the observations.


\begin{table}
\caption{Log of Optical Observations at the Nordic Optical Telescope}             
\centering                          
\begin{tabular}{cccc}        
\hline                
UT$^a$ & $T_{expo}$ (s) & Filter & magnitude$^b$  \\
\hline
\multicolumn{4}{c}{PKS~1502+106}\\
\hline
2008 Aug 13, 21:01:28.0 & 120 & $B_{Bes}$   &   $17.45  \pm 0.09$  \\
2008 Aug 13, 21:05:33.0 & 80   &  $V_{Bes}$      &  $17.05  \pm 0.02$  \\
2008 Aug 13, 21:08:51.0 & 60  &  $R_{Bes}$      &  $16.63  \pm 0.02$  \\
2008 Aug 13, 21:11:50.0 & 60  &  $i_{int}$           &  $16.06  \pm 0.01$    \\
\hline
\multicolumn{4}{c}{Mkn 841}\\
\hline
2008 Aug 13, 20:47:22.0 & 20 &  $B_{Bes}$  & $14.90  \pm  0.12$ \\
2008 Aug 13, 20:50:43.5 & 15 &  $V_{Bes}$    & $14.63  \pm  0.05$ \\
2008 Aug 13, 20:52:35.0 & 10 &  $R_{Bes}$   & $14.34  \pm  0.03$ \\
2008 Aug 13, 20:54:41.0 & 10 &  $i_{int}$      & $14.21  \pm  0.04$  \\
\hline                        
            \noalign{\smallskip}
\multicolumn{4}{l}{$^a$ Mid time of observation}\\
\multicolumn{4}{l}{$^b$ Not corrected for Galactic dust absorption.  Errors are}\\
\multicolumn{4}{l}{ ~~~ standard deviations of the photometric zero point.} \\
\end{tabular}
\end{table}


%
   \begin{figure}
   \centering
   \includegraphics[width=8cm]{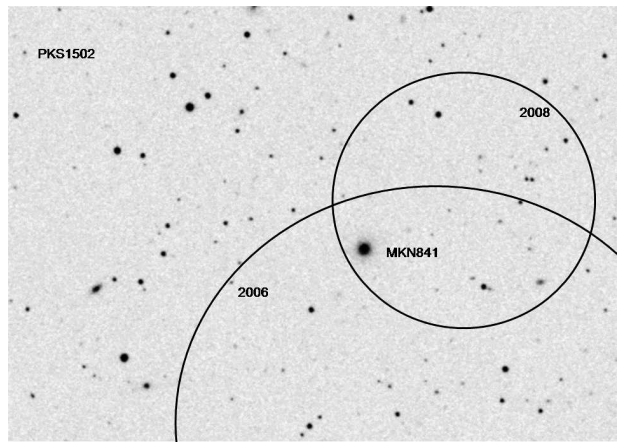}
   \caption{Digitized Sky Survey  image of the field of PKS~1502+106 and Mkn~841 in red filter.  The image vertical size corresponds to 7 arcmin. The 2 optical sources are indicated.  Over-plotted   are the error circles of the IBIS detections of January 2006 (larger circle, with a radius  of 5 arcmin) and August 2008 (smaller circle, radius of 2.4 arcmin)}
              \label{FigImage}
    \end{figure}

%

\section{Results}

\subsection{INTEGRAL}

A source is detected in hard X-rays with IBIS/ISGRI 
at coordinates RA = 15:03:53.5, Dec = +10:26:45, with an error of 2.4 arcmin at 90\% confidence level (see Figure~1).   The detection has an  8.6-$\sigma$ significance in
the co-added image of the full observation.  The centroid of  the error circle of this source is located at 8.3 arcmin from the optical source corresponding to PKS~1502+106, and thus it is inconsistent with the blazar at a 4-$\sigma$ significance level (i.e., the association probability is $\sim 10^{-3}$).

However, it is located at only 1.9 arcmin from Mkn~841 (optical catalogue coordinates J2000: RA = 15:04:01.20, Dec = +10:26:16.2), and therefore consistent with it (see Fig. 1).  
The error circle of the IBIS detection of January 2006 is also reported in Fig. 1: as already noted, this hard X-ray source is also consistent only with the Seyfert galaxy.    

From the count rates measured in our IBIS observation we derive a flux of (2.4 $\pm$ 0.3)
$\times 10^{-11}$ erg~s$^{-1}$~cm$^{-2}$ in the range 15-60 keV for Mkn~841, assuming the spectrum measured by BeppoSAX ($\Gamma  = 2$, Petrucci et al. 2002). The limited signal-to-noise ratio does not allow us to extract a spectrum from the present IBIS observation), and a 3-$\sigma$
upper limit of $9.3 \times 10^{-12}$ erg~s$^{-1}$~cm$^{-2}$ in the same energy range for the
blazar.  These values are similar to the ones determined by Ubertini et al. (2008) from the January
2006 IBIS observation, taking into account the slight difference in energy range.
 
Neither PKS~1502+106 nor Mkn~841 are detected by JEM-X in both  2006 and 2008.    Assuming a Crab-like spectrum, we estimate 3-$\sigma$ upper limits on the 3-10 keV flux of both sources of  $3.3 \times 10^{-11}$ erg~s$^{-1}$~cm$^{-2}$ and  $2.3 \times 10^{-11}$ erg~s$^{-1}$~cm$^{-2}$  for 2006 and 2008, respectively (consistent with the XRT measurements, see below).

The OMC has measured an optical magnitude of  $V = 15.4 \pm 0.4$ for PKS~1502+106.  Since this is substantially brighter than the V-band flux measured by NOT (see Table~1) and 
considering the relatively large OMC pixel size  (about 18 arcsec), we have checked for possible contamination from nearby sources.  Four USNO stars are located within 1 arcmin of the blazar:  the summed contribution of their magnitudes, when added to the flux of the blazar as measured with the NOT,  is comparable to the measured magnitude so that they contaminate significantly the blazar.  Therefore, we will ignore its OMC optical measurement. The OMC field of view included also  Mkn~841, for which we measured an average magnitude of   $V = 14.33 \pm 0.02$  with no significant variability.  No nearby source contaminates significantly the Seyfert optical emission.

\subsection{Fermi LAT}

We have re-analyzed the {\it  Fermi} LAT spectrum  in outburst state presented by Abdo et al. (2010b) using 2 different time intervals for  signal integration: for the same period of five days between 5 and 10 August 2008  adopted by Abdo et al. (2010b) we obtain a photon index of $\Gamma = 2.05 \pm 0.04$, consistent with that measured by those authors ($\Gamma = 2.06 \pm 0.02$). For the day of the gamma-ray flare peak, 8 August 2008, we find a  photon index of  $\Gamma = 1.88 \pm 0.08$.  These indices are reported in Figure~3 (see Discussion).

\subsection{{\it Swift}}

Comparison of our results with those of Abdo et al. (2010b) for PKS~1502+106  and of Grupe et al. (2010) for  Mkn~841  shows a general consistency.   {\it Swift} XRT and UVOT light curves for PKS~1502+106  in August 2008 have been reported by Abdo et al. (2010b), who found a decreasing trend in X-rays, the aftermath of the gamma-ray outburst, and a fully resolved flare in optical/UV,   correlated with the gamma-ray event.  In Figure 2a  we show the light curves of Mkn~841 during the present campaign and in Figure 2b we report all XRT and UVOT V-band flux measurements of the Seyfert galaxy,  to compare its  long- and short-term variability in X-ray and optical.  Significant X-ray variability of up to $\sim$60\% was detected on a time scale of 10 days and of $\sim$50\% in almost 2 years.   The V-band variability is less pronounced on both time scales, only about 15-20\%.     The light curves in the other UVOT filters have a behavior similar to the V-band, with somewhat larger variability amplitude, although the UV filter data have a larger error than the optical ones.   It is interesting to note that the long-term X-ray and optical variations are correlated (Fig. 2b).

The XRT spectrum of the blazar on 8 August 2008 is satisfactorily ($\chi^2 = 32.5$ for 28 d.o.f.) fitted by a single power-law $F_E  \propto E^{-\Gamma}$ with  photon index $\Gamma = 1.52 \pm 0.09$ (errors on fitted X-ray spectral parameters are at  90\% confidence), and the  flux in 2-10 keV is  $1.8 \times 10^{-12}$ erg~s$^{-1}$~cm$^{-2}$.  These results are consistent with those of Abdo et al. (2010b), who integrated the X-ray spectral signal over 4 days around the outburst.  The extrapolation of the XRT  flux to the IBIS energy range of 15-60 keV  yields a flux of $3.8 \times 10^{-12}$ erg~s$^{-1}$~cm$^{-2}$, which is 6 times lower 
than the flux measured by IBIS for the detected source.  This  independently suggests that, barring an improbably large spectral change between the soft and hard X-ray ranges, the  IBIS  source  does not coincide with the blazar. 

A single power-law describes well  ($\chi^2 = 46.3$ for 52 d.o.f.) the spectrum of Mkn~841 of 8 August 2008.  The  photon index is $\Gamma = 2.03 \pm 0.07$  and the fitted flux in 2-10 keV is $1.2 \times 10^{-11}$ erg~s$^{-1}$~cm$^{-2}$.   The extrapolation to the IBIS energy range yields a flux of 
$9.9\times 10^{-12}$ erg~s$^{-1}$~cm$^{-2}$ (15-60 keV), which is 2.4 times lower than measured by IBIS, and therefore  compatible with it, both taking into account the not exact simultaneity of the XRT spectrum and IBIS detection and assuming a spectral change between soft and hard X-ray range due to the emergence of a reflected Compton hump (see Discussion).  No significant  X-ray   spectral  variability is seen during the August 2008 campaign.
Comparing our XRT spectrum of Mkn~841 with those of previous epochs reported by Grupe et al. (2010) we find some modest, marginally significant X-ray  spectral variability, correlated with the flux in the sense of a harder spectrum in lower X-ray emission state.  The broad-band UVOT spectra indicate a rather constant UV-optical spectral slope.

\subsection{Nordic Optical Telescope}

The results of the NOT observations are reported in Table~1.  For the photometry of Mkn~841 we used a rather small aperture (5 pixel radius equivalent to 0.95 arcsec) in order to reduce the host galaxy contribution.  However, we noted that the profile of one of the
comparison star and the target are very similar in $B$- and $V$-band  images, leading us to conclude that the host galaxy contribution is  negligible.   The NOT measurement in $V$-band of Mkn~841 is consistent with the {\it INTEGRAL} OMC measurement within the errors and taking into account the difference in filter central wavelength.

%
   \begin{figure*}
   \vspace{-10cm}
   \centering
   \includegraphics[width=18cm]{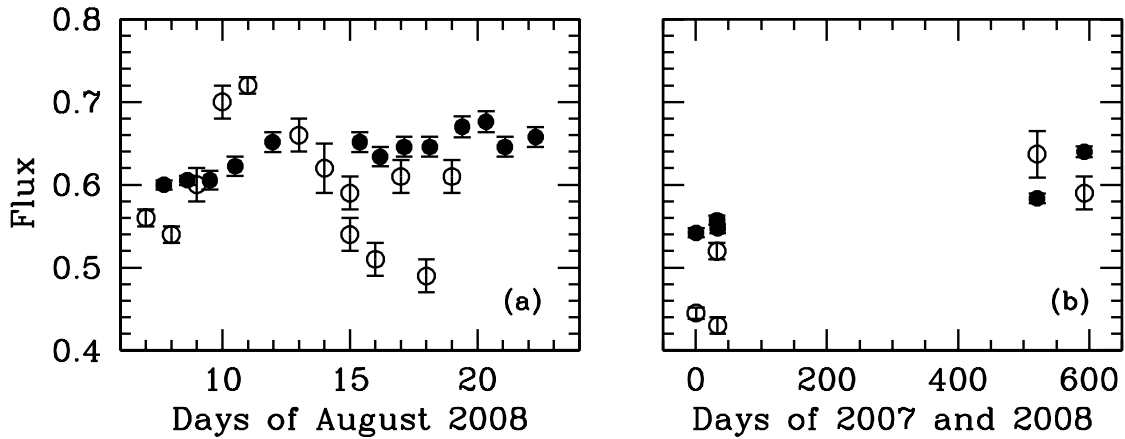}
   \caption{{\it  Swift} light curves of Mkn~841: (a) in  August 2008  and (b) in 2007-2008:   XRT count rates (in c~s$^{-1}$, open circles) and UVOT fluxes (filled circles) obtained with an arbitrary magnitude-to-flux conversion coefficient from the V-band magnitudes.  The errors are statistical only and correspond to 1-$\sigma$.  The data points in (b) are from Grupe et al. (2010), except the last XRT and UVOT ones, that correspond to the averages of the August 2008 points for the two instruments shown in (a).  The zero  time in (b)  corresponds to 2007 January 1, 00 UT}
              \label{FigLC}
    \end{figure*}

%

\section{Discussion}

\subsection{PKS~1502+106}

The data of our multiwavelength campaign in August 2008 are reported in Figure~3, together with  archival non-simultaneous data that sample a quiescent broad-band spectral state.   Also shown is the {\it Fermi} LAT spectrum detected during the first 3 months of the survey (Abdo et al. 2009).
Abdo et al. (2010b) have modeled the SED of  PKS~1502+106 with a synchrotron component at the radio-to-UV wavelengths and inverse Compton scattering at the higher energies.  In particular, synchrotron self-Compton radiation is mostly responsible for the X-ray spectrum, while external Compton upscattering is invoked to reproduce the MeV-GeV spectrum that has a curved shape in the lower state and a power-law slope  during the flare (5-10 August 2010). 

In order to fit both the August 2008 low and high states and the archival SEDs we  applied a simple, one--zone, homogeneous leptonic model based on synchrotron radiation at radio-to-UV frequencies, and inverse Compton upscattering
of both synchrotron photons (self-Compton) and  external photons of the broad emission line region (BLR),
dominating at X-ray and gamma-ray frequencies, respectively, following an approach similar to that of Abdo et al. (2010b).   The details of the model are described in Ghisellini, Tavecchio, \& Ghirlanda  (2009) and Ghisellini et al.  (2010).  
A strong constraint of the model is the variation of the MeV-GeV spectrum in August 2008,  indicating that the peak frequency of the external Compton component changes significantly  from low to outburst state.  Note that the {\it INTEGRAL} IBIS upper limit also  sets a stringent constraint. 
Although the optical-UV spectrum follows a steep single power-law and does not require an additional thermal component (Fig.~3), the presence of an external Compton component at gamma-rays demands a powering source of the BLR.  We identify this with an accretion disk whose luminosity is limited by the smooth shape of the optical-UV spectrum to be no larger than  $L_d \simeq 1.7 \times 10^{46}$ erg~s$^{-1}$, consistent with that postulated by Abdo et al. (2010b).  This is $\sim$11\% of the Eddington luminosity for an assumed central black hole mass of $\sim 10^9 M_\odot$,  therefore in this blazar the jet power exceeds by a large amount the accretion power (see Ghisellini et al. 2010).   The radius of the BLR is assumed to be $R_{\rm BLR} = 4.1\times 10^{17}$ cm.  The total X--ray corona luminosity is assumed to be 30\% of $L_{\rm d}$,
its spectral shape is assumed to be  $\propto \nu^{-1} \exp(-h\nu/150~{\rm keV})$.
The model parameters for the present SEDs are  reported in Table~2 and the corresponding model curves are overplotted to the data in Figure~3 in the same colors indicated in the table.  
The jet viewing angle $\theta_{\rm v}$ is $2.3^\circ$ for all states.

Our SED analysis differs from that of Abdo et al. (2010b) in the changes  of the parameters that are responsible for variability among the various emission states (see Table~2).  At variance  with Abdo et al. (2010b) the parameters that exhibit the largest variations in our modeling are: the magnetic field $B$,  that increases by an order of magnitude between the  highest and lowest gamma-ray state;  the total power injected in the jet $P^{\prime}_i$,  that is a factor 10 higher in the highest than in the lowest gamma-ray state; the random Lorentz factor $\gamma_b$ of the electron distribution cooling break,  that  is significantly larger in the highest gamma-ray state, consistent with the harder spectrum.   The variation of the magnetic field is consistent with the dominance of the inverse Compton cooling in the brightest gamma-ray state of 8 August 2008:  a higher intensity of the  inverse Compton scattering radiation implies that the ratio between the radiation energy density and the magnetic field energy density is relatively large, while the magnetic energy density dominates this ratio in lower inverse Compton states.  The  increase of the injected power $P_i^{\prime}$ is related to an enhancement of the activity of the acceleration mechanism,   so that the observed correlation between $P_i^{\prime}$ and   $\gamma_b$ is not unexpected. We also find that the shape of the particle distribution changes
between the low and the hard states, with harder slopes below $\gamma_b$ and steeper above.
The synchrotron component, traced by the optical data, does not vary as dramatically as the inverse Compton component (MeV-GeV)  in the same time scale, a behavior seen, for instance, in the blazars 3C~279 (see Wehrle et al. 1998 and Giuliani et al. 2009) and PKS~1454-354 (Ghisellini et al. 2009), but different from that of other well monitored blazar sources like 3C~454.3 (Bonning et al. 2009; Vercellone et al. 2009; Donnarumma et al. 2009; Vercellone et al. 2010; Pacciani et al. 2010) and PKS~1510-089 (Abdo et al. 2010e; D'Ammando et al. 2010), where the correlated variations in optical and gamma-rays have comparable amplitudes on time scales from days to years (see however Ghisellini et al. 2007).

A multiwavelength monitoring of few bright {\it  Fermi} blazar sources on a
range of time scales from days to few years would allow us to obtain sensitive insight  into the
undoubtedly complex jet behavior.  If used systematically in these campaigns,  {\it INTEGRAL} will sample, or  will  set constraining limits on, the flux at hard X-ray energies, that is crucial for locating precisely the peak of the high energy component.   Besides this intrinsic merit,  {\it INTEGRAL} IBIS  -- with its large field-of-view  --  often allows serendipitous detections of sources located within a few degrees of the primary target.  Both in the present case, with the detection of the Seyfert Mkn~841, and in other previous circumstances (e.g., Pian et al. 2005),  this  advantage  has led  to an additional contribution in the study of the hard X-ray extragalactic sources.

\subsection{Mkn~841}

{\it BeppoSAX}  observations of  the Seyfert 1 Mkn~841 in 1999 and simultaneous {\it XMM}-Newton and {\it BeppoSAX} observations in 2001 showed a high
energy cut-off at around 100 keV (Bianchi et al. 2001; Petrucci et al. 2002; Bianchi et al. 2004).  As noted by Ubertini et al. (2008), 
this strongly suggests that this source, despite being 20 times brighter
in X-rays than PKS~1502+106 when quiescent, is unlikely to emit in the {\it  Fermi} LAT domain at the reported level (Abdo et al. 2010b).  Therefore, the {\it  Fermi} LAT detection can be firmly identified with emission from the blazar.  On the other hand, the imaging and spectral arguments made in the previous section suggest that the hard X-rays detected by IBIS should be entirely attributed to the Seyfert.     

From the January 2006 and  present {\it INTEGRAL} IBIS  observations  there is no evidence of significant variability of the hard X-ray flux of Mkn~841.  This source was first detected in hard X-rays (50-150 keV) in 1996   by {\it CGRO} OSSE (Zdziarski, Poutanen, \& Johnson 2000), at a flux level of (6 $\pm$ 2) $\times 10^{-11}$ erg~s$^{-1}$~cm$^{-2}$,  an order of magnitude higher than the extrapolation of the IBIS flux in the OSSE band using a  reflected Compton spectrum (see below and Fig.~4), a difference that is not to be considered dramatic owing to  the low significance of the OSSE detection (we tend to exclude contamination of the OSSE detection by PKS~1502+106, see below).  On the other hand, the {\it  Swift} BAT detection of  Mkn~841 obtained during the first 39 months of the mission (Cusumano et al. 2010) is somewhat lower than the IBIS flux (see Figure~4).  This indicates a moderate secular  variability of the hard  X-ray emission.  The softer  (0.3-10 keV)  X-ray  flux varies more substantially  during the {\it  Swift} mission lifetime, and exhibits maximum amplitude variability of about $\sim$50\% on both months and days time scales (Fig.~2), in agreement with the behavior observed by other  satellites, that detected even faster variations (George et al. 1993; Nandra et al. 1995; Bianchi et al. 2001; Petrucci et al. 2007).

In Figure 4 we report  the spectral energy distribution  of Mkn~841 based on the {\it INTEGRAL}, {\it  Swift}, and NOT quasi-simultaneous (within few days) observations  of August 2008, and compare it  with the historical data and with the {\it  Swift} XRT and UVOT observations of February 2007.   The {\it  Swift} BAT  average flux and spectrum from the 39-month survey (Cusumano et al.  2010) are also shown.    The spectrum has the overall typical shape of a Seyfert 1 nucleus, with a prominent rising optical/UV component, a relatively soft  spectrum in the 0.3-10 keV range and a hard X-ray component with a cutoff around 150 keV, that  suggests a relatively high coronal maximum temperature.   The  NOT, UVOT and XRT measurements of August 2008 indicate a relatively high state.   The limited accuracy of XRT does not allow us to appreciate the possible presence of the soft X-ray excess, that is often observed in Seyferts and has been first detected precisely in this object  (Arnaud et al. 1985).  However, as already noted, the lower state of February  2007 is also harder, suggesting a variable soft component, more prominent in the high states.   The correlation between X-ray spectral index and flux (harder X-ray spectrum during lower X-ray states) is confirmed by all {\it Swift} XRT observations of this source (see Grupe et al. 2010).

We have overimposed to the data, without any scaling,  the model of a black body with characteristic temperature of $\sim$100 eV  to reproduce phenomenologically the soft X-ray excess,  plus a  power-law due to Compton upscattering of the coronal soft X-rays, with a cutoff at 100 keV, and a reflected continuum and fluorescent iron line component at the hard X-rays, proposed  by Bianchi et al. (2001) as the best description of the 1999 BeppoSAX spectrum (see also Ponti et al. 2008 for  a Suzaku observation of the hard X-ray spectrum).  Apart from the iron line, that is out of the reach of the limited sensitivity and spectral resolution of XRT, the model reproduces satisfactorily  the XRT, IBIS and BAT data (Fig. 4).   

The average optical-to-X-ray colour index $\alpha_{ox}$  obtained from the present and previous (Grupe et al. 2010) XRT and UVOT observations of Mkn~841 is consistent with the relationship between $\alpha_{ox}$ and UV luminosity found in optically selected AGNs (Strateva et al. 2005; Steffen et al. 2006), although this is not the case when the $\alpha_{ox}$ values of the individual {\it Swift} observations are compared, because the X-ray variations are larger than the optical ones.   Correlated X-ray and optical variability  with larger  amplitude in X-rays than optical has been seen in other Seyferts (e.g., Breedt et al. 2010), although the opposite is also seen in objects of this AGN class. This behavior of Mkn~841 optical-X-ray variability and the correlated X-ray flux and spectral variations (steeper spectrum for brighter states) must be both taken into account when constructing an emission scenario for this source.  However, our  data are not adequate to provide a detailed physical characterization of the soft X-ray excess component and its relation to the optical-UV bump. 
This requires a complex model (see e.g. Petrucci et al. 2007; Ponti et al. 2008; Longinotti et al. 2010) and an accurate and sensitive simultaneous monitoring in optical/UV and X-rays.

We conclude with a note on the issue of  possible contamination of the hard X-ray flux of Mkn~841 in the  observations that preceded the {\it INTEGRAL}  one:  the model we used to reproduce the broad-band spectrum of PKS~1502+106  (Fig.~3) predicts an intensity of about $5 \times 10^{-12}$ erg~s$^{-1}$~cm$^{-2}$ in hard X-rays, with little variability among very different MeV-GeV emission states.  This is a rather negligible fraction (20\%) of the IBIS detected flux.  Assuming therefore that neither the blazar nor the Seyfert vary dramatically in hard X-rays, both the {\it BeppoSAX}  PDS measurement of Mkn~841 (Bianchi et al. 2001; Petrucci et al. 2002), which is consistent with IBIS,  and the  {\it CGRO}  OSSE measurement (Johnson et al. 2000)  do not contain significant contribution from   the blazar, which lies in the field of view of the two instruments during Mkn~841 observations.  The  uncertainty circle  of the {\it Swift} BAT Mkn~841 detection has a radius of 2.5 arcmin (Cusumano et al. 2010), which  excludes contamination from PKS~1502+106.


\begin{table*} 
\centering
\caption{Parameters of the theoretical SEDs of PKS~1502+106. }
\begin{tabular}{lllllllllllll}
\hline
\hline
date$^a$  &$R_{\rm diss}^b$  &$P^{\prime,c}_{\rm i}$  &$B^d$ &$\Gamma^e$ 
   &$\gamma^f_{\rm b}$ &$\gamma^g_{\rm max}$ &$s^h_1$  &$s^i_2$  &$\log P^j_{\rm r}$ &$\log P^k_{\rm B}$  &$\log P^l_{\rm e}$   &$\log P^m_{\rm p}$  \\
(1) & (2) & (3)  & 4) & (5) & (6) & (7) & (8) & (9) & (10) & (11)  & (12) & (13)   \\
\hline   
{\it Fermi} LAT, 8 Aug 2008  (light green)   &300 (1000)   &0.16   &0.75 &19   &3000 & 6000    &--1   &2.6 & 46.8 & 44.8  & 44.9 & 47.2 \\ 
{\it Fermi} LAT, 5-10 Aug 2008  (orange) &195 (650)    &0.1    &2.2  &19   &500 &5000    &--1   &2.4         & 47.1  & 45.4  & 45.0 & 47.2 \\
15-22  Aug 2008 (cyan)           &210 (700)    &0.05   &2.2  &19   &250 &4300  &--1   &2.3                          & 46.3 & 45.5 & 44.8 & 47.1 \\
historical (black)          &120 (400)    &0.015  &7.3  &10   &800 &2000    &0.25  &2.2                                  & 45.1 & 45.5  & 45.5 & 45.9 \\
\hline
            \noalign{\smallskip}
\multicolumn{13}{l}{$^a$ epoch/state (colors refer to the model curves in Fig. ~3)}\\
\multicolumn{13}{l}{$^b$ dissipation radius in units of $10^{15}$ cm and (in parenthesis) in units of gravitational radii}\\
\multicolumn{13}{l}{$^c$ power injected in the blob calculated in the comoving frame, in units of $10^{45}$ erg s$^{-1}$}\\
\multicolumn{13}{l}{$^d$  magnetic field, in Gauss}\\
\multicolumn{13}{l}{$^e$ bulk Lorentz factor at $R_{\rm diss}$}\\
\multicolumn{13}{l}{$^f$  break random Lorentz factors of the injected electrons}\\
\multicolumn{13}{l}{$^g$ maximum   random Lorentz factors of the injected electrons}\\
\multicolumn{13}{l}{$^h$ slope of the injected electron distribution [$Q(\gamma)$] below  $\gamma_{\rm b}$}\\
\multicolumn{13}{l}{$^i$ slope of the injected electron distribution [$Q(\gamma)$] above $\gamma_{\rm b}$}\\
\multicolumn{13}{l}{$^j$ Logarithm of the jet power in the form of radiation, in erg s$^{-1}$}\\
\multicolumn{13}{l}{$^k$ Logarithm of Poynting flux, in erg s$^{-1}$}\\
\multicolumn{13}{l}{$^l$ Logarithm of bulk motion power of electrons, in erg s$^{-1}$}\\
\multicolumn{13}{l}{$^m$ Logarithm of bulk motion power of protons, in erg s$^{-1}$}\\
\end{tabular}
\end{table*}


%
   \begin{figure*}
   \centering
   \includegraphics{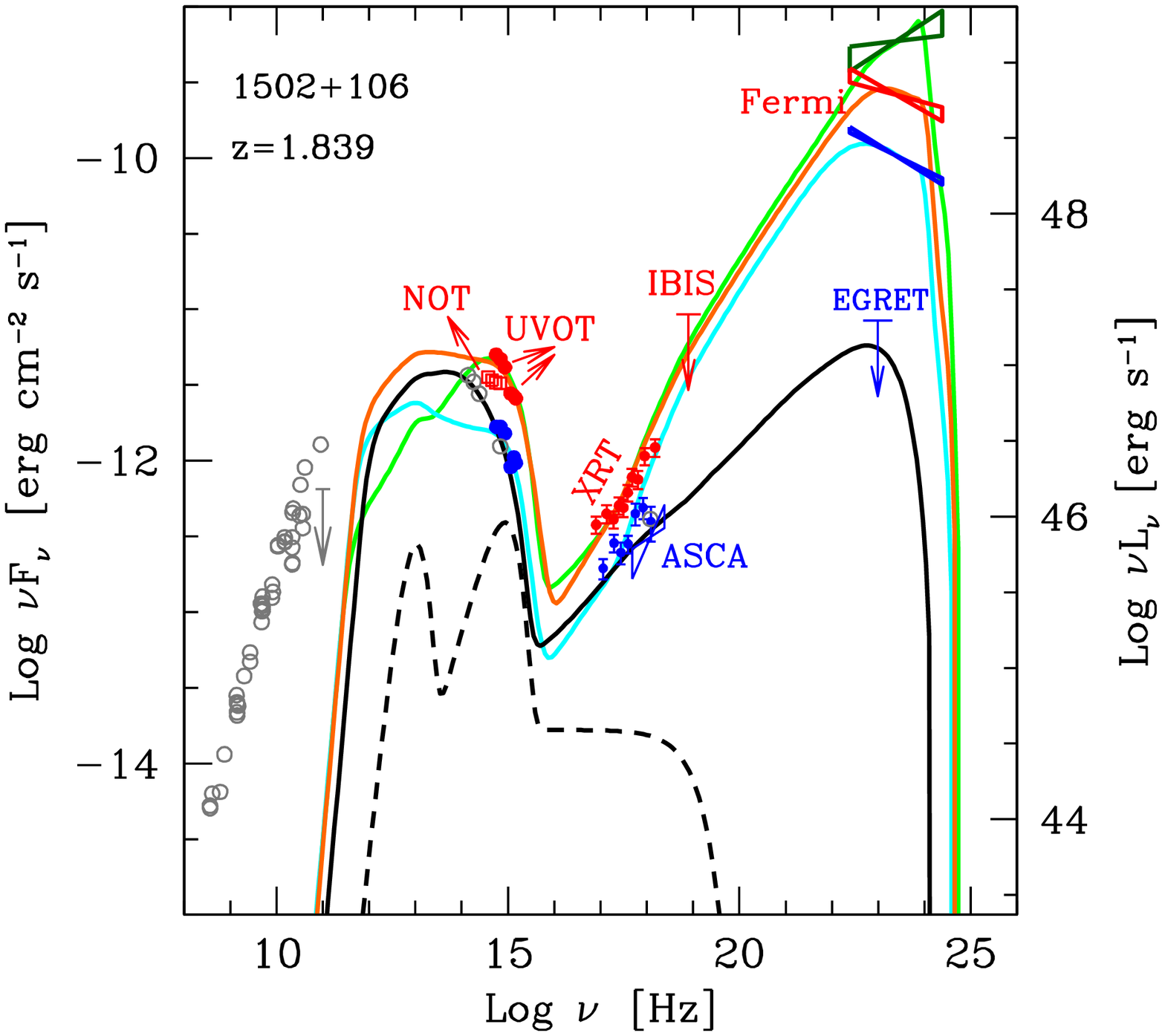}
   \caption{Spectral energy distributions of PKS~1502+106.   The red  circles and the darkgreen bowtie at GeV energies represent the data of the outburst state of 8 August 2008; the red bowtie represents the   {\it  Fermi} LAT outburst spectrum averaged over the days 5-10 August and the downward pointing arrow  is the {\it INTEGRAL} IBIS/ISGRI upper limit of 9-11 August.   The blue circles represent the post-flare state of 15-22 August 2008 ({\it  Swift} XRT and UVOT) and the blue bowtie is the spectrum in the low gamma-ray state observed by {\it  Fermi} during  its first 3 months in orbit (Abdo et al. 2009), equivalent to the {\it  Fermi} LAT post-flare state after 11 August 2008 (see Abdo et al. 2010b).  The NOT data (red open squares) refer to 13 August 2008.  The {\it ASCA} and {\it CGRO} EGRET data (blue)  and the grey open circles represent a historical quiescent radio-to-gamma-ray state. 
The optical/UV  data have been corrected for dust extinction in our Galaxy using $E_{B-V} = 0.032 $ (Schlegel et al. 1998) and the extinction curve of Cardelli et al. (1989).  
The optical magnitudes from NOT and the optical-UV magnitudes from UVOT have been converted to fluxes according to Fukugita, Shimasaku, \& Ichikawa (1995) and to Poole et al. (2008), respectively. 
The errors include the statistical and systematic uncertainties.  The solid curves represent the synchrotron plus synchrotron self-Compton and external Compton model we used to fit the data of the highest gamma-ray state (light green), average outburst gamma-ray state (orange), the low state (cyan) and the historical quiescent state (black).  See text and Table~2 for the model parameters.
The black  dashed curve represents the thermal emission from the dust torus, the accretion disk and its corona.  The models do not account for the radio emission, that is likely produced in larger regions than the homogeneous one where the high energy emission takes place.}
              \label{FigSed1502}
    \end{figure*}

%

%
   \begin{figure*}
   \centering
   \includegraphics{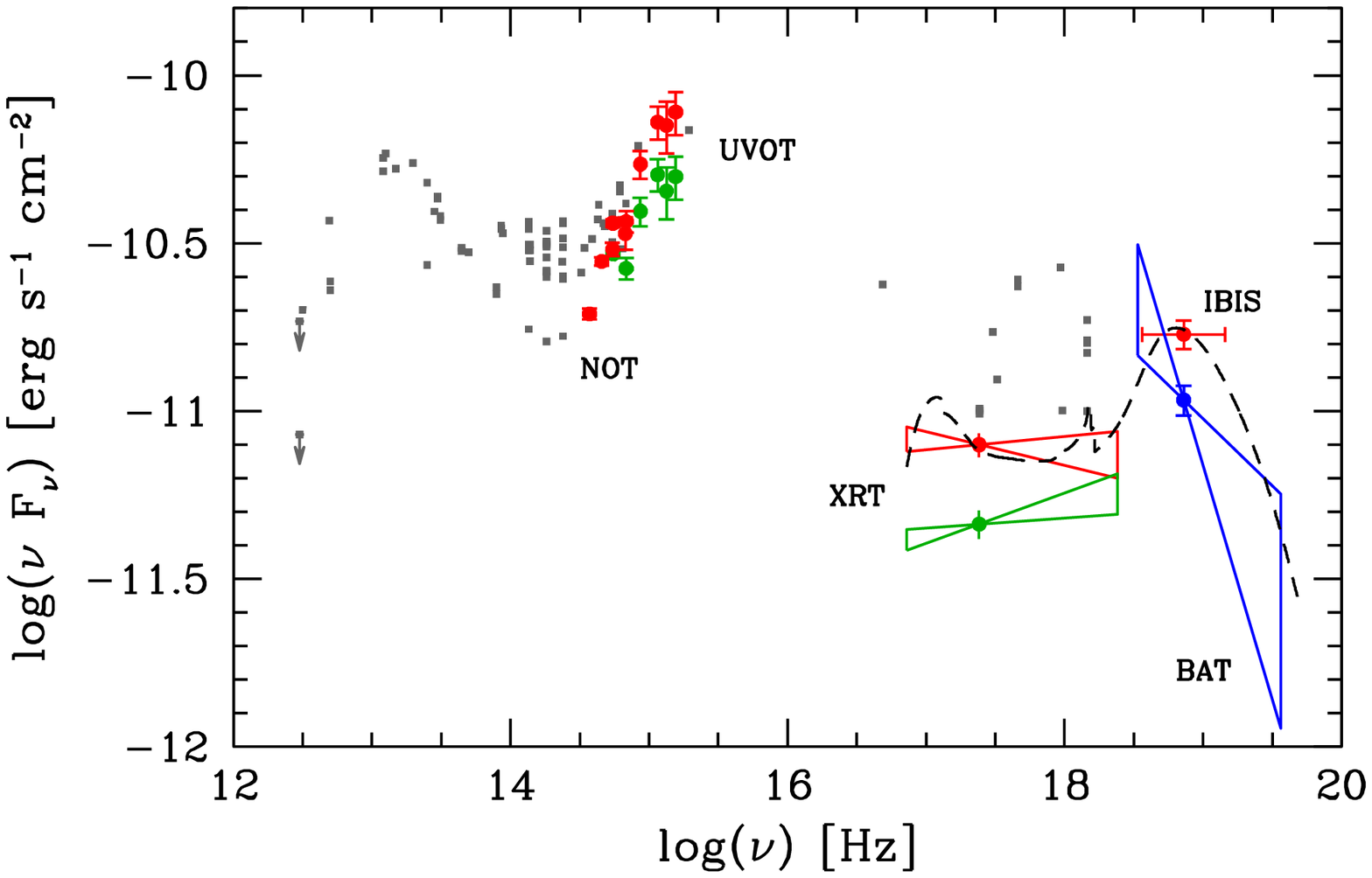}
   \caption{Spectral energy distributions  of Mkn~841 in August 2008 (red symbols: {\it INTEGRAL} IBIS/ISGRI, 9-11 August;  {\it  Swift} XRT spectrum, 8 August;  {\it  Swift} UVOT, 13-15 August; NOT photometry, 13 August), on 3 February 2007 (green symbols:  {\it  Swift} XRT and UVOT) and at prior epochs, not from simultaneous data (taken from  the NED Extragalactic Database, grey squares). The optical/UV data have been corrected for dust extinction in our Galaxy using $E_{B-V} = 0.03 $ (Schlegel et al. 1998) and the extinction curve of Cardelli et al. (1989).  The optical magnitudes from NOT and the optical-UV magnitudes from UVOT have been converted to fluxes according to Fukugita, Shimasaku, \& Ichikawa (1995) and to Poole et al. (2008), respectively.  In addition, the  {\it  Swift} BAT  spectrum from the first 39 months of the mission is shown in blue (December 2004 - February 2008, Cusumano et al.  2010).   The errors include the statistical and systematic uncertainties.    The model fitted by Bianchi et al. (2001) to the 1999 {\it BeppoSAX} spectrum, including a soft X-ray excess, a Comptonized spectrum of a hot corona with a high energy cutoff and a reflected component,  is shown as a dashed curve.}
              \label{FigSed841}
    \end{figure*}

%

\begin{acknowledgements}
We thank C. Winkler, P. Kretschmar and C. Sanchez for assistance with the {\it INTEGRAL} observations scheduling, and P. Lubinski and the {\it INTEGRAL} Science Data Center  staff  for help with data reduction.  We benefitted from  financial support through contracts ASI-INAF I/023/05/0 and ASI-INAF  I/088/06/0. 
RH acknowleges support from ESA PECS 98023 and by the GA CR 102/09/0997 and 205/08/1227.
This research has made use of NASA's Astrophysics Data System Bibliographic Services,  of the SIMBAD Astronomical Database, and of the NASA/IPAC Extragalactic Database (NED), which is operated  by the Jet Propulsion Laboratory, California Institute of Technology, under contract with  the National Aeronautics and Space Administration. 
\end{acknowledgements}

\end{document}